\documentstyle[aps,epsfig,preprint]{revtex}

\newcommand{\be}{\begin{equation}}
\newcommand{\ee}{\end{equation}}
\newcommand{\bea}{\begin{eqnarray}}
\newcommand{\eea}{\end{eqnarray}}
\newcommand{\bsa}{\begin{subeqnarray}}
\newcommand{\esa}{\end{subeqnarray}}

\newcommand{\Veff}%
{{\cal V}^{\mbox{\scriptsize p}}_{\mbox{\scriptsize eff}}}

\begin{document}
\tighten

\title{Interplay of Three-Body Interactions in the EOS of Nuclear Matter}
\author{W. Zuo}
  \address{Institute of Modern Physics, Chinese Academy of Sciences,
          730000 Lanzhou, China}
\author{A. Lejeune}
  \address{Institut de Physique, B5 Sart-Tilman, B-4000 Li\`ege 1, Belgium}
\author{U. Lombardo}
  \address{Dipartimento di Fisica, 57 Corso Italia, and
           INFN-LNS, 44 Via Santa Sofia 95125 Catania, Italy }
\author{J. F. Mathiot}
  \address{Laboratoire de Physique Corpusculaire, Universit\'e Blaise-Pascal,
            CNRS-IN2P3, 24 Avenue des Landais, F-63177 Aubiere Cedex, France }
\maketitle
\begin{abstract}
The equation of state of symmetric nuclear matter has been investigated 
within Brueckner
approach adopting the charge-dependent Argonne $V_{18}$ two-body force plus
a microscopic three-body force based on a meson-exchange model.
The effects on the equation of state of the individual processes giving
rise to the three-body force are explored up to high baryonic density.
It is found that the major role is played by the competition between the
strongly repulsive $(\sigma, \omega)$ exchange term with  virtual
nucleon-antinucleon excitation and the large attractive contribution due
to $(\sigma, \omega)$ exchange with $N^*(1440)$ resonance excitation. 
The net result is a repulsive term which shifts the saturation density 
corresponding to the only two-body force much 
closer to the empirical value, while keeping constant the saturation 
energy per particle.   
The contribution from $(\pi, \rho)$ exchange 3BF is shown to be
attractive and rather small.
The analysis of the separate three-body force contributions allows to
make a comparison with the prediction of Dirac-Brueckner approach which is
supposed to incorporate via the {\it dressed} Dirac spinors
the same virtual nucleon-antinucleon excitations as in the present three-body
force. The numerical results
suggest that the three-body force components missing from the Dirac-Brueckner 
approach are not negligible, especially at high density.
The calculation of the nuclear mean field and the effective mass
shows that the three-body force affects to a limited extent such properties. 


\end{abstract}
\newpage

\section{Introduction}

It is generally accepted that in order to reproduce the empirical saturation
properties of nuclear matter in a non relativistic many-body approach one 
needs to introduce three-body forces (3BF). Calculations with 3BF have been 
performed in non-relativistic Brueckner theory \cite{LEJ,MATH,LEJEU,BAL} and 
in variational approaches\cite{WFF,PANDA,PANDA1}.Moreover
in the relativistic Dirac-Brueckner (DB) approach \cite{MALF,BROCK,KUO}, 
3BF effects  are incorporated via the {\it dressed} Dirac spinors as already 
stressed by Brown et al.~\cite{BROWN}.

The existing versions of 3BF applied for nuclear matter calculations are based 
on either a purely phenomenological model with two parameters adjusted
on the empirical saturation point of nuclear matter~(and/or the $^3H$ binding 
energy) or a microscopic meson-exchange model with nucleonic virtual 
excitations. In the recent years a general method based on the chiral 
effective field theory has been developed to systematically generate 
many-body forces \cite{GLOC}.    

In the study of nuclear matter the phenomenological TBF has been applied both 
in variational calculations \cite{WFF,PANDA,PANDA1} and in Brueckner 
calculations \cite{BAL}, the microscopic one only in Brueckner calculations 
\cite{LEJ,MATH,LEJEU}.

The microscopic 3BF is more appropriate to get a deeper understanding of  
saturation mechanism of nuclear matter and also to extend the study of 
the equation of state (EOS) to high
density. In fact these two aspects of EOS are not
necessarily strictly related each other. Moreover it allows to make contact 
with the 
DB theory, where the saturation mechanism is interpreted as a purely 
relativistic effect. It is true in fact that this effect is coming from the
$\bar N N$ virtual excitations in the scalar $\sigma$-meson exchange
process due to the {\it dressed} Dirac spinors in the nuclear medium.
But it is just one of the possible contributions to the 3BF and one has to 
check whether other contributions are important or not.

While the last ten years much studies have been concentrated  on the
contribution from $\pi$ and $\rho$ mediated 3BF, in this study we want to 
emphasize the role played by $\sigma$ and $\omega$ exchange, and in
particular the one played by the excitation of the Roper ($N^*$(1440)) resonance 
in the medium. This is the main motivation of the present study.

The last years the Brueckner many-body approach has done significant progress 
as concerns the convergence of the hole-line expansion and the validity of
the Brueckner-Hartree-Fock approximation. This goal has been reached by means of
 an
accurate estimate of the three-hole line contributions with both the 
standard and continuous choice for the auxiliary potential \cite{SONG}.    
So it seems to be a suitable basis to extend many-body calculations with
two-body forces (2BF) to higher 
order interactions.

The natural framework for 3BF would be the Bethe-Faddeev equation,
describing the interaction of three nucleons in nuclear matter. But one
may get rid of such a complicate job by a suitable reduction of the 3BF
to an effective 2BF, where the correlation effects with the spectator nucleon
are correctly provided by the Brueckner iterative procedure.

The purpose of this paper is to revisiting the Brueckner theory of
nuclear matter with the 3BF introduced in Ref.~\cite{MATH} performing new 
calculations where the 2BF is the updated charge-dependent Argonne $V_{18}$, 
which reproduces the experimental phase-shifts up to high energy \cite{SCHIA}. 
So we may numerically investigate the role played for the nuclear matter EOS 
by the different processes giving rise to 3BF and the limitations inherent the
relativistic DB approach. Such processes dominate the high density behavior of 
nuclear matter, which is a basic input in the studies on the evolution
of astrophysical compact objects. Moreover the 3BF effects on the nuclear
mean field are also discussed in view of a proper description of the
single particle motion in the dynamical simulations of central events of 
heavy-ion collisions (HIC).

\section{Microscopic Three-Body Force}

The microscopic three-body force adopted in the present calculation
is constructed from the meson-exchange current approach as described in
Ref.~\cite{MATH}.
We refer the reader to this reference for a comprehensive 
discussion of the model. Let us however recall here its main properties.

In the present understanding of the microscopic structure of nuclei, the NN
potential originates from the exchange of mesons. In the OBEP approximation,
$\pi, \rho, \sigma$ and $\omega$ mesons are considered. Once the NN potential is
determined, i.e. once the meson-nucleon couplings, form factors and meson masses
are fixed, many body forces are uniquely fixed as well.
The first candidate to 3BF is the well known Fujita-Miyazawa contribution
involving a $\Delta$ resonance in the intermediate state with $2\pi$ exchange 
\cite{FUJ}. 
Extension of this contribution to $\rho$ exchange, together with all other 
$\pi$ and $\rho$ exchange contributions have been intensively studied in the 
past, leading for instance to the Tucson-Melbourne (TM) 3BF \cite{ELLI}. 
For the present study, we stick to the parametrisation of the TM 3BF,
and change only the parameters according to our choice of the NN potential 
(see below). In terms of meson exchange diagrams, the contributions are shown 
in Fig.~1a.

More recent studies \cite{GLOC} have been concentrated on the general
structure of the $2\pi$ 3BF within the framework of chiral perturbation theory.
While these studies are important in order to settle the $2\pi$-3BF on first
principles according to the symmetry properties of QCD, we do not think they can
yet be extended to quantitative investigations of nuclear matter at densities 
between one to three times normal nuclear matter density. In this domain indeed
, the momenta of the exchanged mesons can be large, and the contribution of 
heavy mesons ($\sigma$ and $\omega$) are dominating over the $2\pi$-3BF 
\cite{MATH}. The 3BF associated to these mesons are shown in Figs.~1b and 1d. 
They are discussed in more details in Ref.~\cite{MATH}.
The $\rho\pi$ and $\rho\sigma$ diagrams, plotted in Fig.~1c, belong to the 
meson-meson coupling terms which should give a smaller contribution 
\cite{MATH} and will not be considered here.
More complicated processes could contribute to the 3BF including other 
nucleonic excitations, but one may expect that the lowest energy excitations,
i.e. the isobar $\Delta(1232)$ resonance and the Roper $N^*(1440)$, play the 
major role in the density domain here considered.     

The necessary coupling constants $\sigma NN^*$ and
$\omega N N^*$ are calculated within the relativistic color dielectric model
\cite{CHAN}. The form factors at these vertices are also calculated in this 
model. Note that because of the orthogonality of the radial wave function of 
the $N$ and $N^*$, these form factors cannot have the usual dipole 
parametrization (as chosen in Ref.~\cite{COON} for instance). 
According to Ref.~\cite{MATH} (see Sec.~V) they are parametrized as follows: 
\bea
F_{MNN^*}(\vec{k}^2)=\frac{\Lambda^2+\alpha \vec{k}^2}{\Lambda^2-\alpha m^2}
\left(\frac{\Lambda^2-m^2}{\Lambda^2+\vec{k}^2}\right)^2
\nonumber
\eea
The values of the parameters are reported below ( see Table I).

Once the dynamical origin of the NN potential is determined from one boson
exchange (OBE) mechanisms, all the parameters needed to determine the
3BF are extracted from the NN potential itself. In the case this NN potential 
is directly parametrized in terms of OBE potentials, like the Bonn
potentials, these parameters are given directly from the fit to NN
observables, or NN phase shifts. Instead, in the case it is given
in terms of a purely numerical parametrization, like the Paris
potential or the $AV_{18}$ potential for instance, one has to extract
the equivalent OBE parameters by a direct comparison between various
spin-isospin potential in the OBE form and in the parametrized form.
This procedure was used in Ref.~\cite{MATH} in order to extract meson
exchange parameters from the Paris potential. The $\pi$ and $\rho$
exchange parameters were determined from the $S=1$ tensor potential
(keeping some of the parameters, like masses and coupling constants,
to their physical values), while $\sigma$ and $\omega$ exchange were
determined from the central and spin-orbit potentials.
In the present calculations, the  parameters of the 3BF
have been re-determined from the  OBE potential
model to fulfill the self-consistency with the adopted
two-body interaction $AV_{18}$. It is found that
for the $\sigma$ and $\omega$ exchange, the parameters derived
from the $AV_{18}$ force remain the same as those given for the
Paris potential~\cite{MATH}.
While for the $\pi$ and $\rho$ exchange, the parameters have
to be slightly adjusted in order to fit reasonably both the tensor
part and the spin-spin part of the $AV_{18}$ potential.
The new parameters of the 3BF are collected in Tab.~1.
As for the $\sigma$-meson mass, the value of $540$MeV is still
adopted, which has been checked to satisfactorily reproduce the
$AV_{18}$ interaction from OBE potential.
Varying of the $\sigma$-meson mass from
 $540 MeV$ to $600 MeV$ does not change significantly the
self-consistency with $AV_{18}$, so it is also an interesting
point to investigate the effect of the $\sigma$-meson mass on the 
nuclear matter EOS, which is underway and will be presented elsewhere.

Our procedure to determine the parameters of OBE models from a NN
potential given in a purely numerical form is of course not unique,
nor exact.  In particular, the short range part of the potentials is
not of the same shape because of the ad-hoc regularization used in the
$AV_{18}$ or Paris potentials as compared to the use of form factors
in OBE models.  These parts of the potential are however of no
importance in the contribution of the 3BF to the binding energy of
nuclear matter since there are completely washed out by short range
correlations (see next section). This procedure assures however that
the OBE models used to determine 3BF have a strength and a range for
each meson exchange in agreement with the two-body potential.

\section{ Brueckner approach with 3BF}

The effect of the 3BF is included in the self-consistent Brueckner
procedure along the same line as in~\cite{LEJ,MATH}, where an effective
two-body interaction is constructed without solving the full three-body
problem (for a detailed description and justification of the method we 
refer to Ref.~\cite{LEJ,MATH}). Starting from the 3BF interaction discussed 
in Sec.~II one defines the effective 2BF as 
\bea
  <\vec r_1 \vec r_2| V_3 | \vec r_1 ' \vec r_2 ' > =
  \frac{1}{4} Tr\sum_n \int {\rm d}{\vec r_3} {\rm d} {\vec r_3 '}
\phi^*_n(\vec r_3 ')(1-\eta(r_{13}' ))
(1-\eta(r_{23}'))
\nonumber
\eea
\bea
\times
W_3(\vec r_1^{\ \prime},\vec r_2^{\ \prime}
\vec r_3^{\ \prime}|\vec r_1 \vec r_2 \vec r_3)
\phi_n(r_3)
(1-\eta(r_{13}))(1-\eta(r_{23}))
\eea
where the trace is taken with respect to spin and isospin of the third
nucleon. This is nothing else than the average of the three-body force over 
the wave function of the third particle taking into account via the defect
function $\eta(r)$ the correlations between this particle and the two others. 
The dependence on the defect function entails a selfconsistent determination of
the effective 2BF along with the $G-matrix$ and auxiliary potential in that
we must re-calculate the effective 2BF at each iterative step and then add it
to the bare 2BF for the next loop.

The $\eta(r)$ is actually the defect function averaged over spin and 
momenta in the Fermi sea and it incorporates, for the sake of simplicity,
only the most important partial l-wave components, i.e., the $^1S_0$ and 
$^3S_1$ partial waves. Corrections due to higher angular momenta in the
defect function are expected to be sizeable at high density. They will be
included in forthcoming calculations. 

The reduction of the 3BF to an effective 2BF is justified by the fact that 
three-body correlations are small \cite{SONG}, but ultimately one would solve 
the Bethe-Faddeev equations with 3BF to investigate their effect on 
the rate of convergence of the Brueckner-Bethe-Goldstone expansion.    

\section{ Numerical results }

The procedure followed in our calculations is the same as in the usual
selfconsistent BHF scheme: The $G$-matrix is 
calculated selfconsistently along with the auxiliary potential by solving the 
Bethe-Goldstone equation \cite{BAAL}. Due to its dependence on the defect function the 
effective 2BF is calculated selfconsistently at one step of the iterative 
BHF procedure as said before.

The continuous choice is adopted for the auxiliary potential as it has been 
already well established that it makes the convergence of the hole-line expansion  
much faster than the gap choice~\cite{SONG}. It is calculated in a momentum 
range with cutoff $k_c = 5 fm^{-1}$. 
As bare NN interaction we used the 
Argonne $AV_{18}$ which gives an excellent fit to NN scattering data as well as 
to the deuteron binding energy~\cite{SCHIA}. The partial wave expansion is 
truncated to $L_{max} = 6$, but the effects of higher partial waves, specially
at high density, are to be estimated with further calculations.

\subsection{$\sigma,\omega -\bar N N$  3BF}

This subsection is devoted to the discussion of 
the contribution due to the 2$\sigma$ exchange diagram
with intermediate virtual excitation of a nucleon-antinucleon ($N\bar N$)
pair (hereafter denoted by $2\sigma -(N\bar{N})$ 3BF). This contribution is
worth of particular attention as it is believed to be the main relativistic 
effect present in DB approach~\cite{BROWN,SEROT}.

In order to consider separately the effect of the $2\sigma -(N\bar N)$ 3BF,
the calculations have been done by switching off in the self-consistent 
Brueckner iterative scheme the other 3BF components, i.e., adding only the 
$2\sigma -(N\bar N)$ part to $AV_{18}$. The same procedure has been adopted
when calculating the other individual 3BF contributions.

In Fig.~2 the energy shift from the calculation with only bare 2BF due to 
the $2\sigma -(N\bar N)$ 3BF is plotted as a function of the nuclear matter 
density. As expected it gives a repulsive contribution monotonically 
increasing with density and approximately fulfills the following power law 
\bea
\Delta E/A \simeq 2.4 \cdot (\rho/\rho_0)^{8/3} (MeV). 
\eea
When the nucleon-nucleon correlations (ladder diagrams) are switched off 
the repulsion is enhanced since correlations prevent nucleons from getting 
closer and experiencing a stronger 3BF. The energy shift due to the 
$2\sigma -(N\bar N)$ 3BF was also calculated in this approximation 
(indicated by $\eta=0$ in the figure) and the results are fit by the same 
power law but with the prefactor of Eq.~(2) enhanced 
to 3.6  which is close to the value of 4.2 obtained in the crude estimate of 
Ref.~\cite{BROWN}.

In the same figure is also reported the relativistic correction to the energy
from DB approach, which is defined as the difference between the
full DB calculation and its non-relativistic limit, obtained replacing 
in the DB context the dressed nucleon mass by the bare one~\cite{BROCK}. 
In spite of the
different Bonn potentials the shift is almost the same,
implying that  it is rather insensitive to the two-body interaction
adopted.
Since the nuclear EOS of DB approach could be quite different from
the non relativistic Brueckner approximation,
the only meaningful comparison is in fact
between the $2\sigma -(N\bar N)$ contribution and the previous DB correction.
Around the saturation density they almost overlap
each other and differ at most by $26\%$ at the highest density. This
difference is to be attributed to higher order relativistic effects which
become only significant at high density. It is clear from the comparison that
the most important relativistic correction to the EOS of nuclear matter
in DB approach can be reproduced fairly well by BHF calculation
including only the 3BF originated from the
scalar coupling to the virtual excitation of a nucleon-antinucleon pair.
So the accuracy of the DB prediction to the EOS of nuclear matter depends, 
especially at high density, on to what the extent the other three-body forces 
cancel each other, and in particular the diagrams involving the excitation of 
nucleon resonances like the Roper resonance.

In order to push forward the comparison with DB predictions, in Fig.~2 
(right side) is also displayed the full energy per particle in different 
approximations.  
The $2\sigma -(N\bar N)$ term provides a noticeable
improvement of the saturation density in agreement with the DB results
(Bonn B) as compared to the BHF approximation using pure two-body force.
The agreement between the two calculations is far beyond what we might
expect, the high density deviation being attributed to the defect function  
reducing the $2\sigma -(N\bar N)$ 3BF in our calculation.
In the same figure is also plotted the result of a complete BHF calculation
including all the three-body forces of Fig.~1. Comparing with the BHF EOS 
it is seen that the full three-body force gives an
increasingly strong repulsion to the energy per nucleon in nuclear matter
as increasing density. This repulsion is really able to drive the saturation
point towards its empirical one.
However, looking at the complete Brueckner
calculation, i.e. with the full 3BF, one may easily anticipate the importance
that processes other than the $2\sigma -(N\bar N)$ have to shift the
saturation energy of about $2 MeV$ towards the empirical value. 

Besides the scalar $\sigma$ meson and the nucleon-antinucleon virtual
excitations, vector $\omega$ meson will also contribute to
3BF. In Fig.~3 are depicted the energy corrections of the different
3BF components from $(\sigma, \omega)-(N\bar N)$ (left panel) and the
corresponding nuclear EOS (right panel). While the $2\omega-(N\bar N)$
3BF adds only a comparatively small attractive contribution, the
$\sigma\omega-(N\bar N)$ component gives a strong repulsion which is
even larger than the $2\sigma-(N\bar N)$ one. Consequently the
corresponding EOS turns out to be too stiff and the nuclear matter is
underbound at the saturation point with a very low saturation density.   
So we may conclude that relativistic effects associated to the virtual
$N\bar N$ excitations cannot reproduce the empirical saturation properties. 
The 3BF's from nucleon resonances are expected to compensate this too strong 
repulsion. 

\subsection{Effects of 3BF with nucleon resonances and full EOS}

In Fig.~4 (left panel) are shown the energy corrections of the 3BF components 
from nucleon resonances represented in Fig.~1  and the corresponding
nuclear EOS's (right panel). The effect of
the $\sigma, \omega$-Roper  3BF turns out to be quite sizeable. As
already done in Ref.~\cite{MATH}, we include also for consistency the
contribution from Roper-Anti-Roper excitations. This contribution
is however small, and correct by about 10 \% the contribution from the
Roper resonance. Their net contribution is attractive and comparable in 
magnitude to the repulsion from the $\sigma\omega$-$(N\bar N)$ 3BF as shown 
in the same figure for comparison. 

In Fig.~4 is also reported the $\pi,\rho$ exchange 3BF, which provides
only a relatively small attractive contribution to the energy per
nucleon even at high density in agreement with Ref.~\cite{MATH} and it
has a little effect on the nuclear EOS.

The competition between all these 3BF's from different sources determines 
the final EOS of nuclear matter which saturates around $\rho\simeq 
0.198$fm$^{-3}$ with energy per nucleon $E\simeq -15.08$MeV much better than 
the corresponding value $\rho\simeq0.265$fm$^{-3}$ from the prediction
adopting pure $AV_{18}$ two-body force. As compared to the
DB approach, the present EOS is much softer at high density.
At the corresponding equilibrium density the obtained incompressibility $K$ 
is about $207$ MeV, in agreement with both the empirical value
of $210\pm30$ MeV~\cite{BLAIZ}. This nice property should be
attributed to the effect coming from the excitation of the Roper
resonance, as compared to the strongly repulsive and
density-dependent effect coming from $N\bar N$ excitations. We recall
that this latter effect is the only one present in DB calculations.
Despite the BHF EOS exhibits a flatter curvature (see Fig.~2, right side)
the compression modulus at the minimum is larger \cite{BAL1} as an effect
of the rather large saturation density.

\section{Single particle properties}

The nuclear mean field is one of the basic inputs in the dynamical simulations
of HIC. Transverse flows and other collective observables
have been proven to be very sensitive to whether a mean field is taken from
a stiff or soft EOS or whether it is given a momentum dependence  or not.
In central events of HIC high baryonic densities are reached and 3BF effects
could become sizeable. This is the main motivation for calculating the mean 
field and related quantities in our BHF approximation with 3BF.   

In the left part of Fig.~5 is shown the 3BF effect on the mean field for two
densities $\rho=0.17$ and $0.34 fm^{-3}$.
As expected, the 3BF adds a repulsive contribution to the mean field
when increasing the density. The momentum dependence is not much affected 
because the present 
3BF does not introduce any non local effects. This property can be also seen   
considering the effective mass derived by the slope of the
mean field as a function of momentum. It is defined as
\bea
\frac{m^*}{m} = \frac{k}{m}\left( \frac{{\rm d}e(k)}{{\rm d}k}
\right)^{-1},
\eea
where $e(k)$ is the single particle energy, i.e.,
$e(k) = k^2/2m + U(k)$ determined self-consistently
within the Brueckner approach. The calculated effective mass at
the Fermi momentum $k_F$ is depicted
in the right part of Fig.~5. It is seen that $m^*(k_F)$ decreases as
increasing density for both cases with and without the 3BF.
At relatively low density, the 3BF force almost has no effect
upon $m^*(k_F)$, while at high density
($\rho \gtrsim 0.2$fm$^{-3}$), the 3BF has the effect to
saturate $m^*(k_F)$ around a value of 0.72.  
Therefore the 3BF weakens the depth of the mean field, but does not
significantly change the effective mass (around $k_F$) even at high density
(notice the enlarged scale in Fig.~5).   

Both calculations
with and without the 3BF predict a larger value of $m^*(k_F)$ at
high density than
the DB approach ($m^*(k_F)=0.4$ at $\rho=0.54$fm$^{-3}$) and
its non-relativistic
limit ($m^*(k_F)=0.449$ at $\rho=0.54$fm$^{-3}$)~\cite{BROCK}.
This discrepancy might be related to the
`reference spectrum approximation' adopted in the DB
approach, which assumes a weak momentum dependence of
the relativistic scalar self-energy component and
in the non-relativistic limit corresponds to a constant
effective mass approximation~\cite{BROCK}.
It has been shown in Ref.~\cite{SEHN} that the above momentum dependence
could be strikingly strong for the Bonn potentials.

Another interesting aspect concerning the single particle properties is 
the {\it rearrangement} contribution $M_2$ to the mean
field in the hole-line expansion of the mass operator.
This term rises from the density functional derivative of the energy per
particle in the Brueckner-Hartree-Fock approximation. It brings a sizeable 
contribution to the Hugenholtz-Van Hove theorem \cite{HVH}, which
controls order by order the consistency of the hole-line 
expansion~\cite{HM,ZUOW}.
The 3BF effect on the real part  $U_2$ of the {\it rearrangement} term $M_2$, 
shown in the left side of Fig.~6, is a reduction of at most $6\%$ in the low 
momentum range. The most sizeable contribution of TBF to the mean field is
coming from the functional derivative of three-body energy per particle.
The calculation of this contribution requires some additional investigation.

\section{Overview and conclusions}

We have calculated the EOS of symmetric nuclear matter with a microscopic
3BF including the processes depicted in Fig.~1, which are believed to be
the most relevant ones. The charge-dependent Argonne $V_{18}$ has been chosen
as two-body interaction. The main effect of the microscopic 3BF is to provide
that extra-repulsion which, on the one hand, pushes the saturation density 
towards the empirical value, but, on the other hand, is able to keep a  
satisfactory value for the saturation binding energy . This is
a desirable feature since the empirical saturation energy is already well
reproduced by only 2BF, as shown in Fig.~7.
In the same figure the results with phenomenological
3BF~\cite{BAL,PANDA,PANDA1} are plotted
for comparison. The latter gives an EOS comparable with the present one
but only in the context of the Brueckner formalism \cite{BAL}.
The prediction from the variational approach exhibits a strong discrepancy
with the two others in spite of the rather good agreement at the level of
two-body force (Argonne $V_{18}$). Otherwise the good agreement with our 
results show that the 
phenomenological 3BF works quite well in spite of being so simple (only 
two adjustable parameters). But its applicability is restricted to only the
saturation region, since no constraints can be imposed at higher density.

The effects of the three-body forces from different sources on
the nuclear EOS have been
investigated and their relative importance has been discussed along
with the comparison to the DB approach. It is seen that the
$2\sigma-(N\bar N)$ 3BF alone is able to simulate the most important
relativistic correction to the energy per nucleon in the DB approach.
The latter  can be in fact attributed to a suppression of the attractive
$\sigma$ exchange two-body force felt by a Dirac {\it dressed} spinor
in nuclear medium, which corresponds simply to the $2\sigma-(N\bar N)$
diagram of Fig.~1d. However the present investigation shows that besides
this correction  the effects of the 3BF's from other origins
could also be very large (for example, those from $\sigma\omega$
exchange and the Roper resonance), and their net contribution
is attractive and comparable in magnitude to that of
the $2\sigma-(N\bar N)$ 3BF.
As a consequence the repulsion due to the $2\sigma-(N\bar N)$ 3BF
suffers a reduction of about $35\%$ at high density
and the EOS turns  much softer as compared to both the predictions
of the DB \cite{MALF,BROCK} and relativistic mean field approaches 
\cite{SEROT}.

Finally, the Brueckner-Hartree-Fock mean field and effective mass have been
discussed in the present context. The momentum dependence is not so much 
affected by the 3BF so that the effective mass at the Fermi momentum is 
quite insensitive to the 3BF up to high density. The effects of 3BF on the 
rearrangement contribution to the mean field and to the effective mass 
are quite modest as well. 

\vskip 1cm
{\bf Acknowledgments:} \\

   One of us (W.~Z.) would like to thank INFN-LNS and
   the Physics Department of the Catania University, Italy,
   for the hospitality extended to him during the time this work was 
   completed.

   This work has been supported in part by the Chinese Academy of Science,
   within the {\it one Hundred Person Project}, the Major State Basic
   Research Development Program of China under No.~G2000077400.

\newpage

\newpage
\begin{figure}
\caption{ Diagrams of the microscopic 3BF adopted for
the present calculations. This 3BF model is proposed in Ref.~[2]
based on the meson-exchange approach.}
\end{figure}

\begin{figure}
\caption{Left side:  $2\sigma-(N\bar N)$ 3BF contribution to the
energy shift from the 2BF energy per particle in symmetric nuclear matter 
as a function of density.
The squares are the results of BHF calculation including
nucleon-nucleon correlations and the solid curve
denotes simply a fit. The circles
are the corresponding BHF results without nucleon-nucleon
correlations (i.e., setting $\eta = 0$ in Eq.~(1). The dashed curve is a fit.
For comparison are also plotted the relativistic corrections
in DBHF approach, taken from Ref.~[9].
Different symbols distinguish different
version of the Bonn potential. Dotted lines are drawn to guide eyes.
Right side: EOS of symmetric nuclear matter.
The long-dashed curve is the BHF prediction including all 3BF diagrams
(Fig.~1). The solid curve corresponds the result using only
the $2\sigma$ exchange 3BF from
the virtual excitation of a $N\bar N$ pair.
The short-dash one is that using pure $AV_{18}$.
The present results are compared to the DBHF prediction using Bonn B
potential (dot-dash curve, from Ref.~[9]). }
\end{figure}

\begin{figure}
\caption{ 
Left side: Individual $(\sigma, \omega)$--$(N \bar N)$ 3BF contributions 
to the energy shift vs. density 
Right side:
EOS with $(\sigma, \omega)$--$(N \bar N)$ 3BF's in comparison to the EOS from 
both the full 3BF and only pure 2BF}  
\end{figure}

\begin{figure}
\caption{Left side: Individual contributions to the energy shift due the 3BF 
from
nucleon resonance excitations and from $\pi$ and $\rho$ exchange. 
For comparison are also
plotted the full 3BF and relativistic energy shifts.  
The dotted lines are the kinetic energy per particle $E_{k}/A$ and the two body
potential energy per particle $E_{V}/A$.
Right side:
EOS from the left side 3BF components and from pure 2BF in comparison to the
full calculation. 
}
\end{figure}

\begin{figure}
\caption{ Effects of the 3BF on the mean field vs momentum for a couple of 
densities (left panel) and the effective mass calculated at $k=k_F$ vs density 
(right panel). The symbols are the real data, whereas the lines are drawn 
to guide eyes.
}
\end{figure}

\begin{figure}
\caption{ 3BF effects on the ground state
correlations. The {\it rearrangement} term (real part) in the left side and
corresponding effective mass in the right side are plotted vs. momentum 
for two densities.} 
\end{figure}

\begin{figure}
\caption{ Comparison of the present nuclear EOS with the BHF prediction
adopting the phenomenological Urbana 3BF and the variational
calculation using the Urbana $IX$ 3BF.}
\end{figure}
\newpage
{\footnotesize{Tab.~I - Coupling constants and
form factors adopted in the 3BF
consistent to the $AV_{18}$ two-body interaction, which are slightly
different from that given for the Paris potential (see Ref.[2]).
The meson masses
are $m_{\pi}=138$MeV, $m_{\omega} = 780$MeV, $m_{\rho}=776$MeV, and
$m_{\sigma}=540$MeV.}
\\
\begin{center}
\begin{tabular}{ c c c c } \hline\hline
   & ~~~~~$g^2/4\pi$~~~~~ & ~~~~~$\Lambda$ (MeV/$c$)~~~~~
   & ~~~~~$\alpha$~~~          \\ \hline
  $\sigma NN$ & 11.9 & 1100 &  \\
  $\omega NN$ & 33.0 & 1300 &  \\
  $\pi    NN$ & 14.4 & 1580 &  \\
  $\rho   NN$ & 0.55 & 1300 &  \\
  $\sigma NN^*$ & 2.58 & 1450 & $-2.35$ \\
  $\omega NN^*$ & 4.13 & 1550 & $-2.33$ \\
\hline \hline
\end{tabular}
\end{center}
\vskip 1.cm
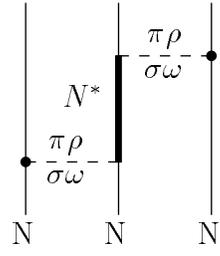
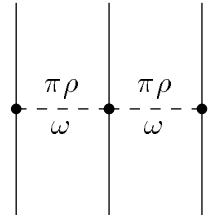
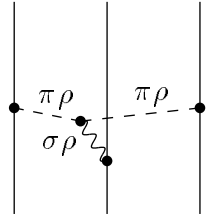
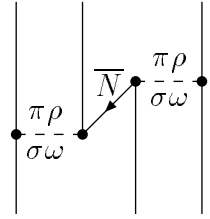
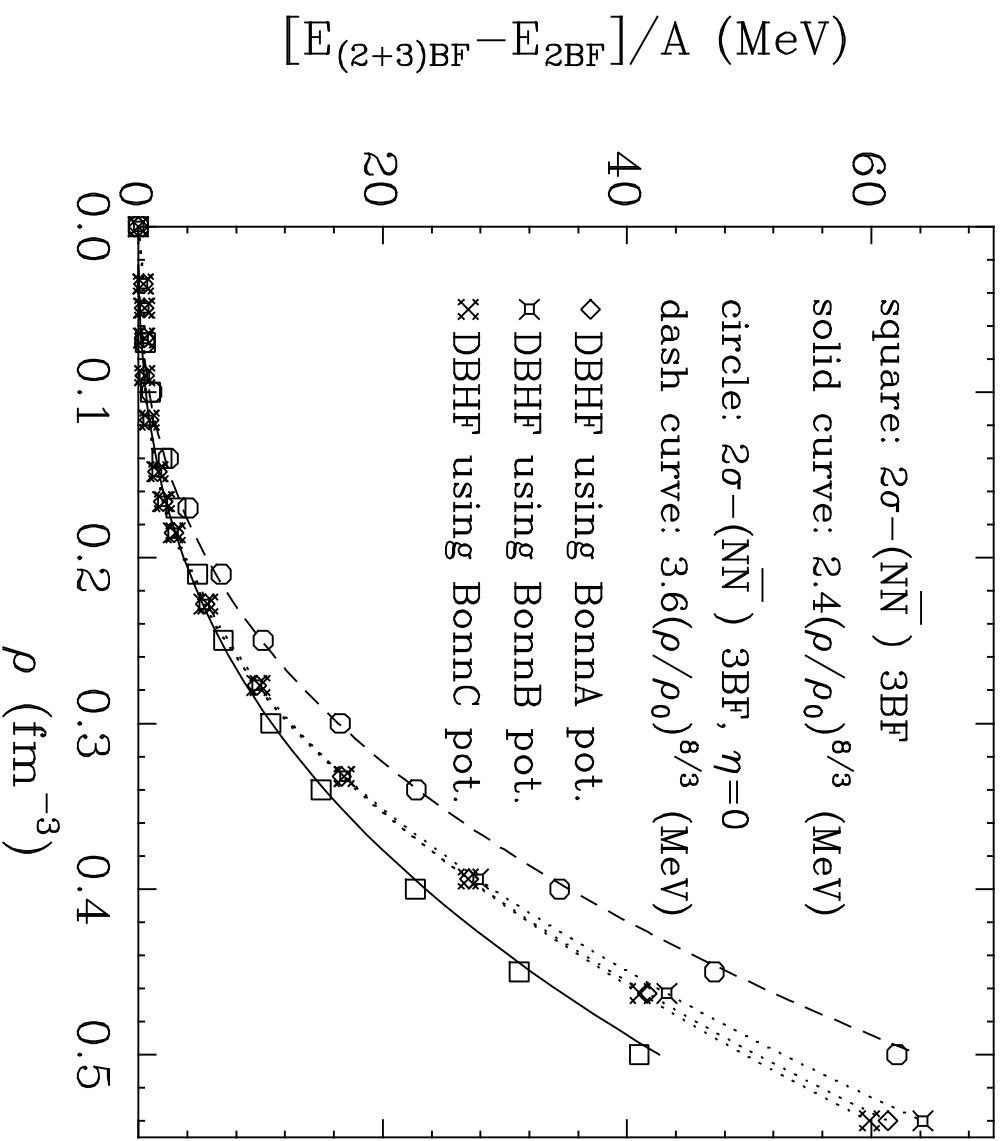
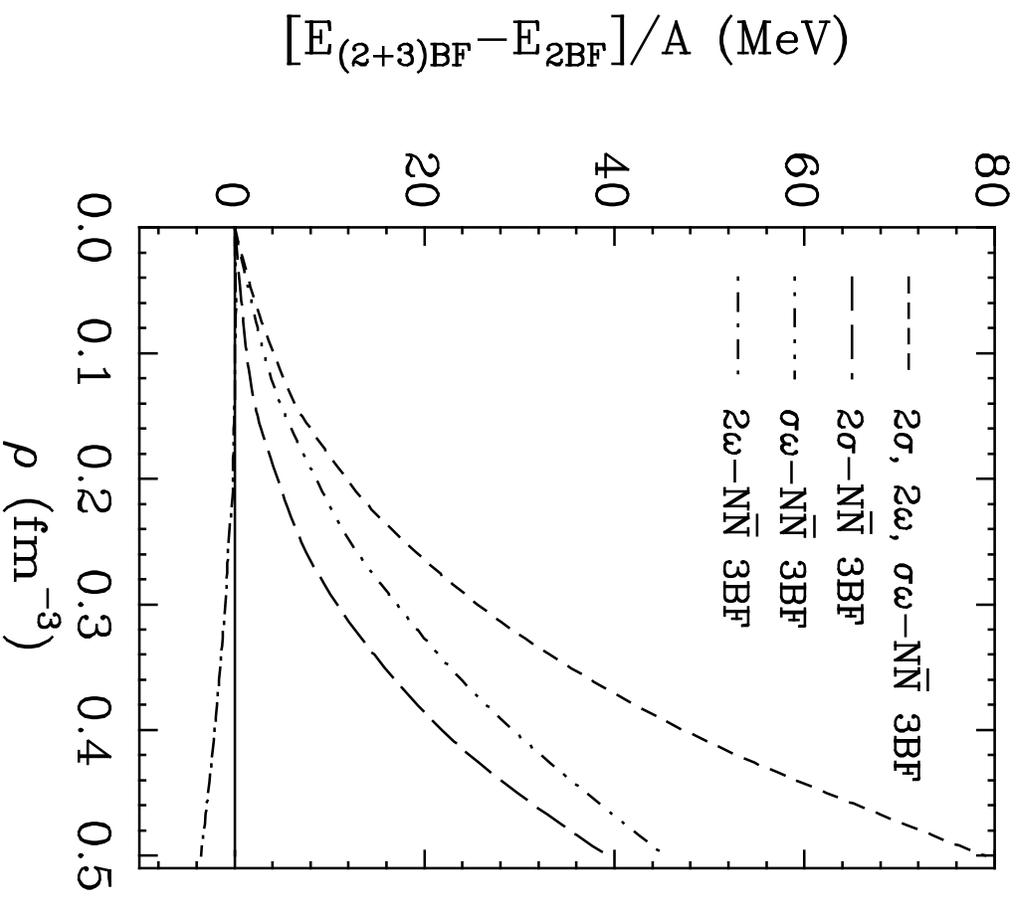
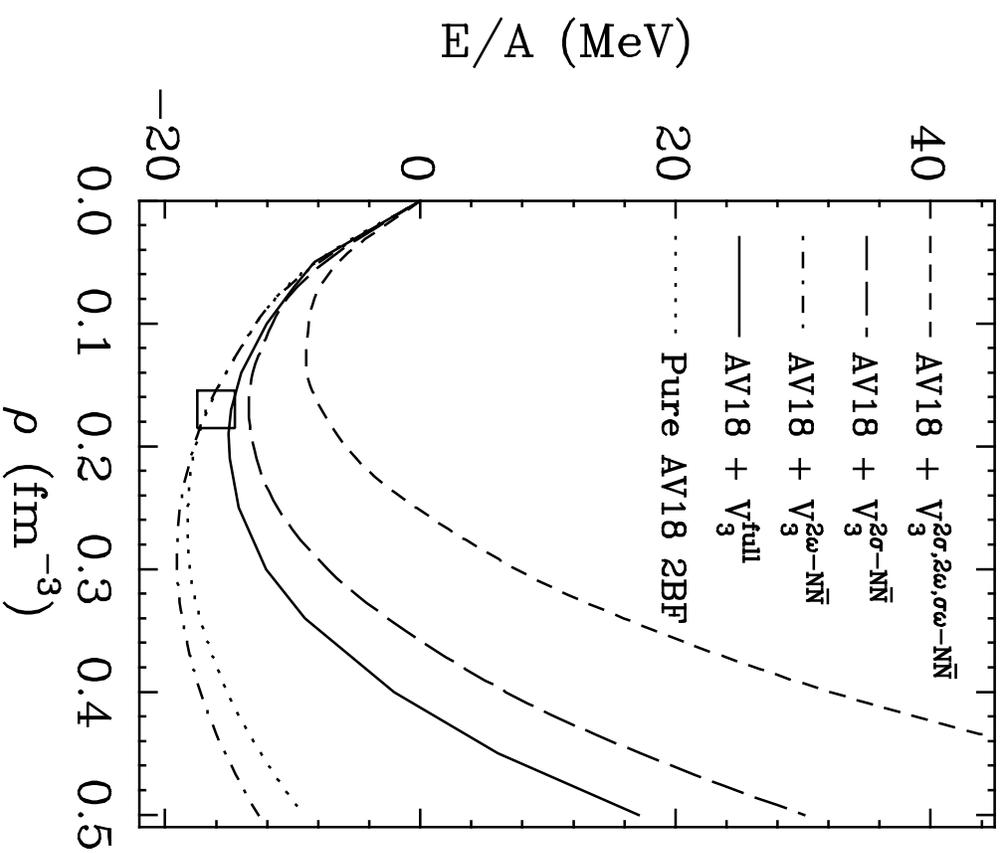
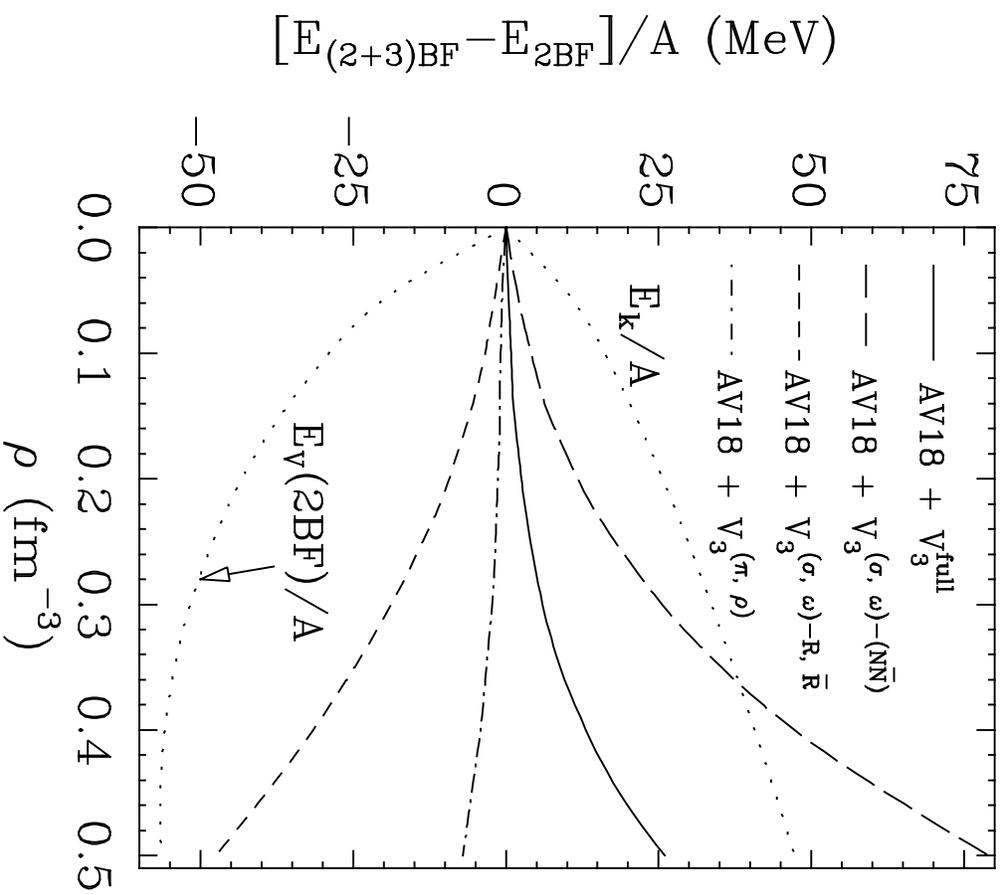
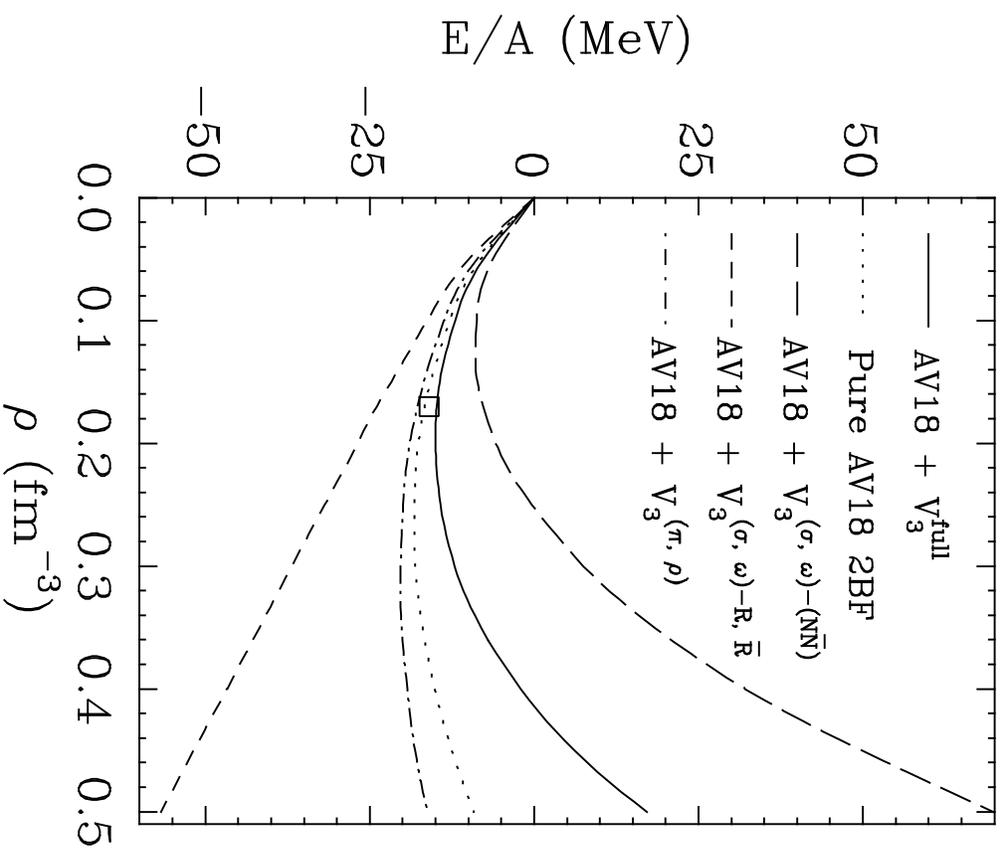
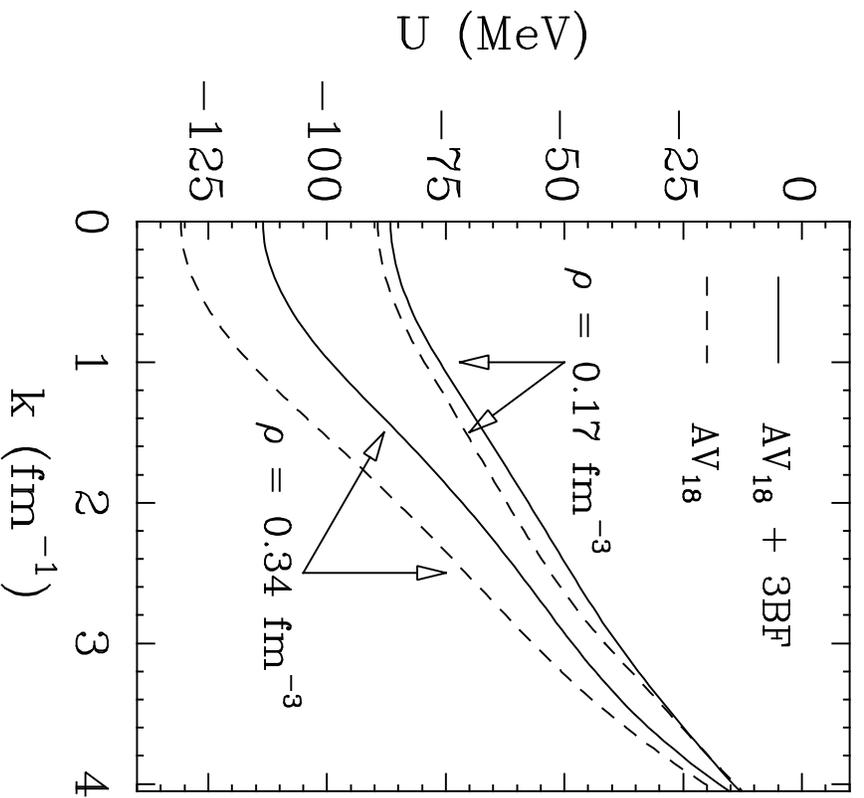
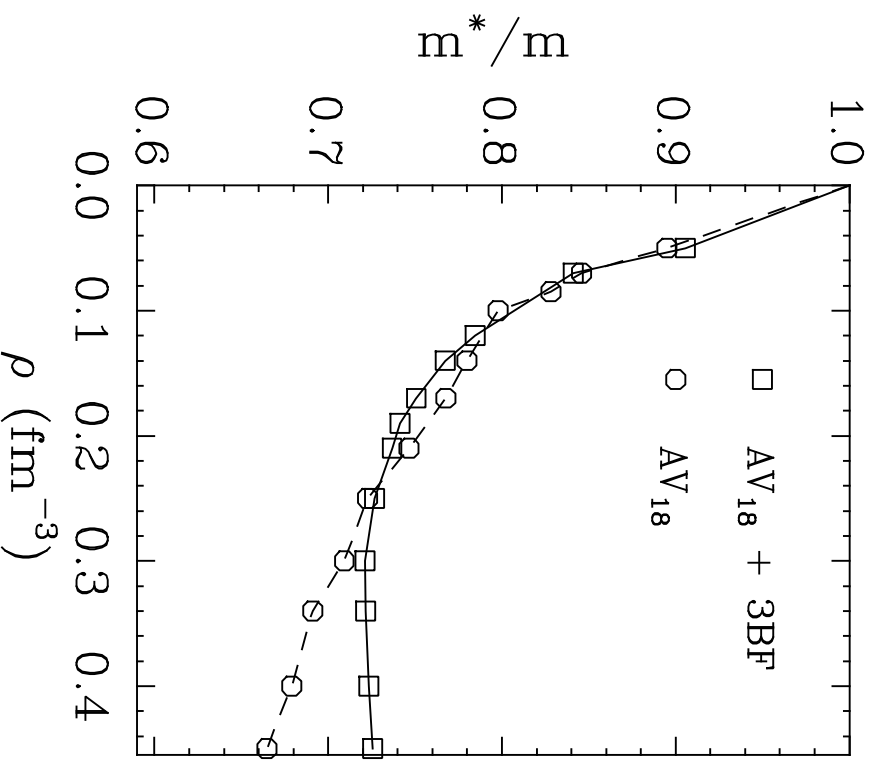
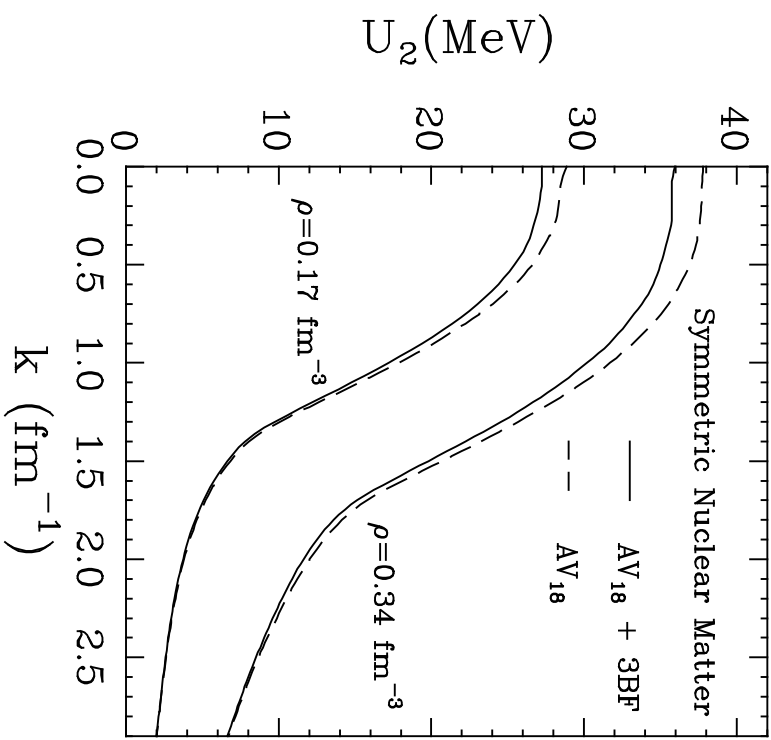
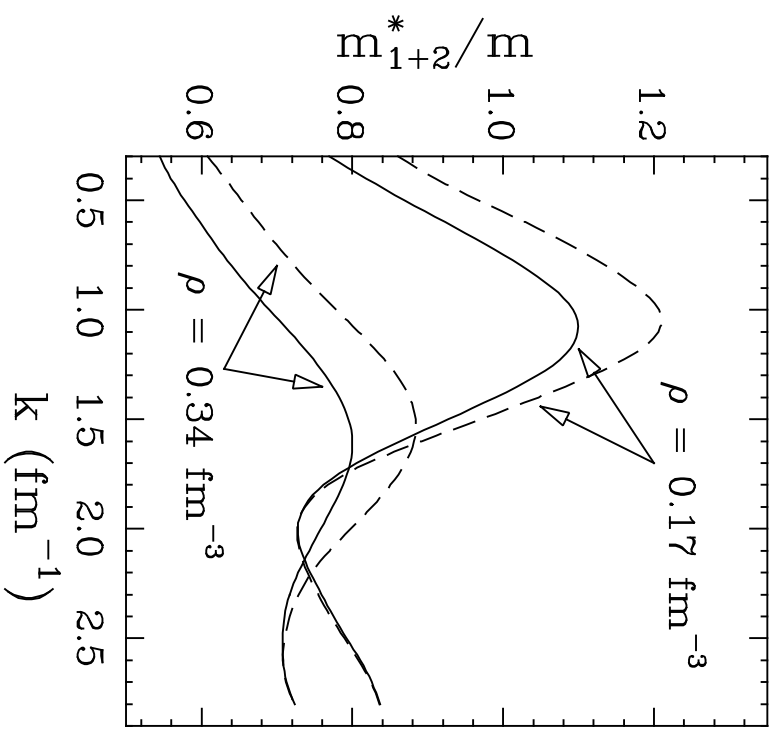
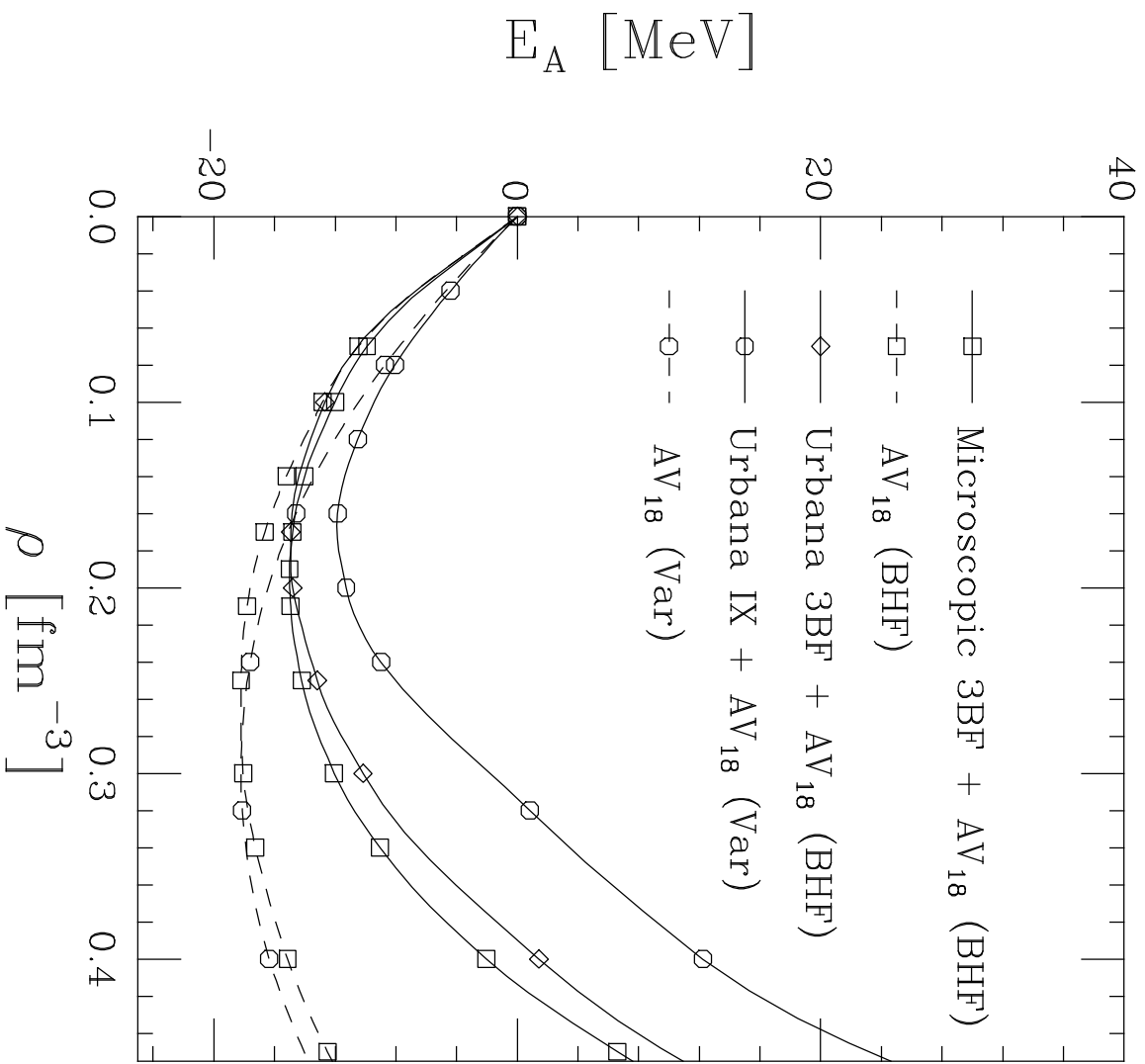
\end{document}